\documentclass[12pt]{article}
 \usepackage{epsfig}
 \def\be{\begin{equation}}
 \def\ee{\end{equation}}
 \def\bea{\begin{eqnarray}}
 \def\eea{\end{eqnarray}}
 \usepackage{graphicx}

 \catcode`\@=11
 \def\lsim{\mathrel{\mathpalette\@versim<}}
 \def\gsim{\mathrel{\mathpalette\@versim>}}
 \def\@versim#1#2{\vcenter{\offinterlineskip
 \ialign{$\m@th#1\hfil##\hfil$\crcr#2\crcr\sim\crcr } }}
 \catcode`\@=12

 \catcode`@=12
\begin{document}
 \thispagestyle{empty}
 \begin{flushright}
 UCRHEP-T582\\
 August 2017\
 \end{flushright}
 \vspace{0.6in}
 \begin{center}
 {\LARGE \bf Radiative Left-Right Dirac Neutrino Mass\\}
 \vspace{1.2in}
 {\bf Ernest Ma$^1$ and Utpal Sarkar$^2$\\}
 \vspace{0.2in}
{\sl $^1$ Physics \& Astronomy Department and Graduate Division,\\
 University of California, Riverside, California 92521, USA\\}
\vspace{0.1in}
{\sl $^2$ Physics Department, Indian Institute of Technology,\\
Kharagpur 721302, India\\}
 \end{center}
 \vspace{1.2in}

\begin{abstract}\
We consider the conventional left-right gauge extension of the standard
model of quarks and leptons without a scalar bidoublet.  
We study systematically how one-loop radiative Dirac neutrino masses 
may be obtained.  In addition to two well-known cases from almost 30 years 
ago, we find two new scenarios with verifiable predictions.
\end{abstract}

 \newpage
 \baselineskip 24pt

\noindent \underline{\it Introduction}~:~\\
To explain why neutrino masses are so small, one approach is to consider the 
case where they are forbidden at tree level and only arise radiatively.  In 
the standard model (SM) without a singlet right-handed neutrino, the 
left-handed neutrino may only acquire a mass through a dimension-five 
operator~\cite{w79}, i.e.
\begin{equation}
{\cal L}_{M} = - \frac{\kappa_{ij}}{\Lambda}(\nu_i \phi^0 - l_i \phi^+) 
(\nu_j \phi^0 - l_j \phi^+) + H.c.
\end{equation}
This means that neutrino masses are Majorana and suppressed by the large scale 
$\Lambda$.  In 1998, it was shown~\cite{m98} how this operator may be 
realized in three and only three ways at tree level (establishing thus the 
nomenclature of Types I, II, and III seesaw), as well as in one loop, 
assuming only fermions and scalars in the loop.  Whereas Majorana neutrino 
masses are theoretically desirable in this context, there is at present 
still no supporting experimental evidence from neutrinoless double beta decay. 
Perhaps neutrinos are Dirac particles after all, and lepton number is 
actually conserved, in which case the question 
remains as to why they are so small.  A general study in the context of the 
SM has recently appeared~\cite{mp17}.   

In this paper we focus on the conventional left-right gauge model, which 
contains the right-handed neutrino in an $SU(2)_R$ doublet and is thus a 
natural framework for considering Dirac neutrinos.
To connect the $SU(2)_L$ fermion doublet with the $SU(2)_R$ fermion doublet,
a scalar bidoublet is required. Suppose a bidoublet is absent~\cite{bms03}, 
then there are no fermion masses in the minimal model.  However, they may be 
generated by dimenion-five operators, in analogy with Eq.~(1).  Specifically, 
Dirac neutrino masses come from the operator
\begin{equation}
{\cal L}_{D} = - \frac{\kappa_{ij}}{\Lambda}(\bar{\nu}_{iL} \bar{\phi}_L^0 
- \bar{l}_{iL} \phi_L^-) (\nu_{jR} \phi_R^0 - l_{jR} \phi_R^+) + H.c.
\end{equation}
These operators may be realized at tree level using heavy singlet quarks and 
leptons, i.e. the mechanism of Dirac seesaw.  Suppose only quark and 
charged-lepton masses are obtained in this fashion.  Neutrinos would then 
appear to be massless.   However Eq.~(2) may still be realized in one loop 
and neutrinos acquire small Dirac masses, as detailed below.

\noindent \underline{\it Four Generic Structures}~:~\\
There are four and only four generic structures which realize Eq.~(2), 
in exact analogy to how Eq.~(1) is realized in Ref.~\cite{m98}. 
The idea is very simple.  To connect $\nu_L$ with $\nu_R$ in one loop, 
we need an internal fermion line and an internal scalar line.  There 
are thus only four ways to do this. 
\begin{itemize}
\item{(A) Attach both $\phi_L$ and $\phi_R$ to the fermion line.}
\item{(B) Attach both $\phi_L$ and $\phi_R$ to the scalar line.}
\item{(C) Attach $\phi_L$ to the fermion line and $\phi_R$ to the scalar line.}
\item{(D) Attach $\phi_L$ to the scalar line and $\phi_R$ to the fermion line.}
\end{itemize}

\noindent \underline{\it Model (A)}~:~\\
A possible implementation of this idea is to add a charged scalar singlet 
$\chi^-$, as shown in Fig.~1.  
\begin{figure}[htb]
\vspace*{-2.5cm}
\hspace*{-3cm}
\includegraphics[scale=1.0]{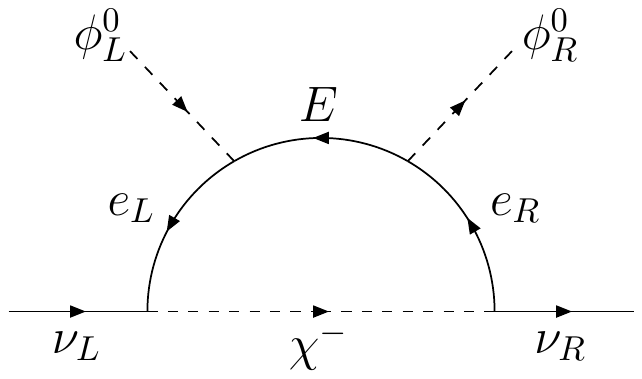}
\vspace*{-23.0cm}
\caption{Dirac neutrino mass in Model (A).}
\end{figure}
This was done many years ago~\cite{m88} and 
it implies that charged leptons have seesaw masses from three heavy singlet 
leptons $E$.  It is the left-right analog of the Zee model~\cite{z80}, but 
without the need for a second scalar doublet.  Note that the $d$ quark 
may be used instead of $e$, then $\chi^-$ should be replaced with a colored 
scalar singlet with charge $-1/3$.

\noindent \underline{\it Model (B)}~:~\\
To make the connection in this case, the heavy singlet lepton is again used, 
as shown in Fig.~2.  This was also done many years ago~\cite{m89}.  
\begin{figure}[htb]
\vspace*{-2.5cm}
\hspace*{-3cm}
\includegraphics[scale=1.0]{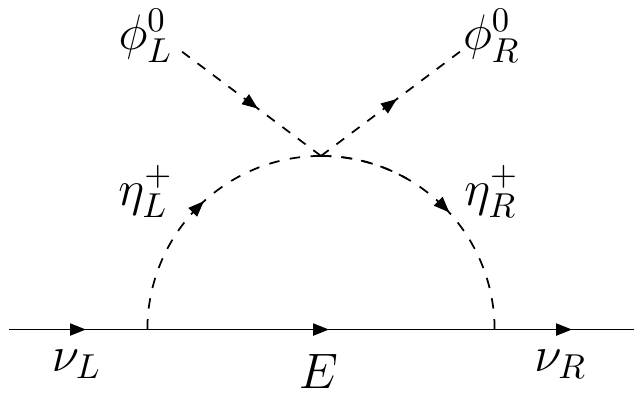}
\vspace*{-23.0cm}
\caption{Dirac neutrino mass in Model (B).}
\end{figure}
A second scalar $SU(2)_L$ doublet $(\eta_L^+,\eta_L^0)$ as well as a 
second scalar $SU(2)_R$ doublet $(\eta_R^+,\eta_R^0)$ 
are needed, because the invariant quartic scalar coupling is required 
to be of the form
\begin{equation}
(\phi_L^+ \eta_L^0 - \phi_L^0 \eta_L^+)^*(\phi_R^+ \eta_R^0 - 
\phi_R^0 \eta_R^+).
\end{equation}
It is thus also a left-right analog of the Zee model, but 
without the charged scalar singlet.  Without loss of generality, we 
have chosen in Fig.~2 $\langle \eta^0_{L,R} \rangle = 0$, so that 
$\eta^\pm_{L,R}$ are the physical charged scalars, whereas $\phi^\pm_{L,R}$ 
have become the longitudinal components of $W^\pm_{L,R}$.

If the heavy $E$ lepton is replaced 
by the heavy $D$ quark, then $\eta_L$ and $\eta_R$ are replaced by the 
corresponding scalar leptoquark doublets.  In this case, charged leptons 
also obtain radiative masses from these same leptoquark doublets through 
the heavy $U$ quarks.

If we impose a discrete $Z_2$ symmetry such that $\eta_{L,R}$ and $E$ are 
odd, then this is a left-right analog of the well-known scotogenic 
model~\cite{m06} of radiative seesaw neutrino mass through dark matter, 
as discussed recently~\cite{bd17}.

\noindent \underline{\it Model (C)}~:~\\
This is a new proposal and requires the existence of an exotic $SU(2)_R$ 
scalar doublet $(\zeta_R^{++},\zeta_R^+)$.
\begin{figure}[htb]
\vspace*{-2.5cm}
\hspace*{-3cm}
\includegraphics[scale=1.0]{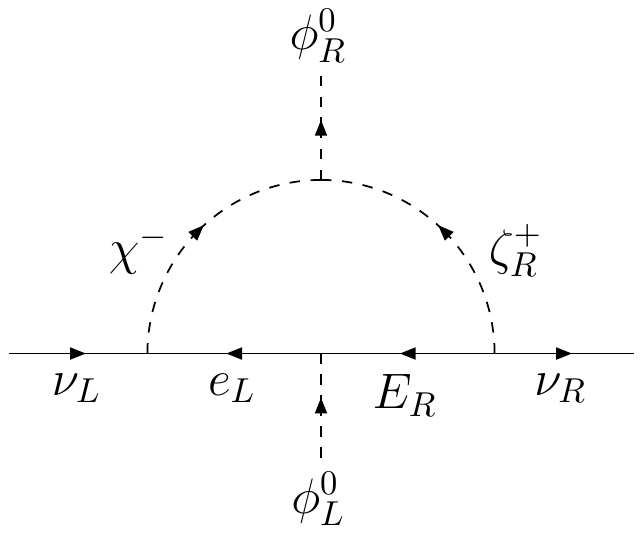}
\vspace*{-22.0cm}
\caption{Dirac neutrino mass in Model (C).}
\end{figure}
Again, if $d$ and $D$ are used instead of $e$ and $E$, $\chi$ and $\zeta$ 
are replaced with the corresponding singlet and doublet scalar leptoquarks 
respectively.

\noindent \underline{\it Model (D)}~:~\\
This is the companion to (C) and requires the existence of an exotic $SU(2)_L$ 
scalar doublet $(\zeta_L^{++},\zeta_L^+)$.
\begin{figure}[htb]
\vspace*{-2.5cm}
\hspace*{-3cm}
\includegraphics[scale=1.0]{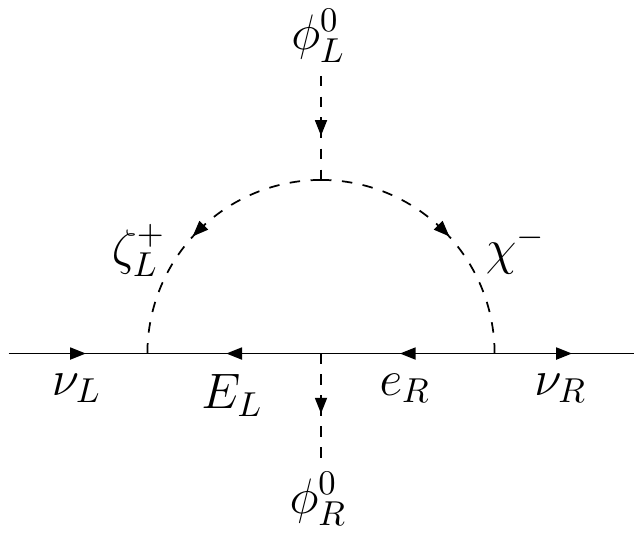}
\vspace*{-22.0cm}
\caption{Dirac neutrino mass in Model (D).}
\end{figure}
Note that if we have both $\zeta_L$ and $\zeta_R$, then Model (B) may be 
realized in Fig.~2 by reversing the arrows of the internal lines and 
replacing $\eta_{L,R}$ with $\zeta_{L,R}$. 

\newpage
\noindent \underline{\it Discussion}~:~\\
To discover which of the above mechanisms is truly responsible for the 
radiative generation of Dirac neutrino masses, the corresponding new 
particles in the loop would have to be observed experimentally. 
Of particular interest are the doubly charged scalars $\zeta_{L,R}^{++}$. 
It is of course well-known that a scalar triplet $(\xi^{++},\xi^+,\xi^0)$ 
may couple to the doublet neutrinos directly and provide them with 
Majorana masses with a small $\langle \xi^0 \rangle$.  In that 
case~\cite{ms98}, the decays of $\xi^{++}$ would map out~\cite{mrs00} 
the elements of the $3 \times 3$ neutrino mass matrix.  These decays 
have been searched for at the Large Hadron Collider (LHC).  Assuming 
100$\%$(50$\%$) branching fraction to $e_L^-e_L^-$, the ATLAS Collaboration 
has the preliminary~\cite{atlas16} lower bound of 570(530) GeV on its mass, 
based on an integrated luminosity of 13.9 fb$^{-1}$ at 13 TeV. However, 
if $\xi$ is indeed responsible for the meutrino mass matrix, its branching 
fraction to $e_L^- e_L^-$ is proportional to the $ee$ entry of the 
$3 \times 3$ Majorana neutrino mass matrix and if that is zero, there 
will be no bound from the LHC in this mode.  Note also that if the decay 
is to $e_R^-e_R^-$ instead, the ATLAS bound becomes 420(380) GeV.
Recently, it was shown~\cite{bj17} that the pair production of doubly 
charged scalars may be enhanced by photon-photon fusion, resulting in 
an improvement of the above limits.

In the models we are concerned with, the doubly charged scalar $\zeta^{++}_L$ 
is part of an $SU(2)_L$ doublet, whereas $\zeta^{++}_R$ is part of an $SU(2)_R$ 
doublet, so it is an $SU(2)_L$ singlet.  This is important because any 
$SU(2)_L$ doublet or triplet will contribute to the $S,T,U$ oblique 
parameters in precision electroweak measurements, but not a singlet. 
In the triplet Higgs model, these are important constraints~\cite{ds16}.
In this context, we note that the new particles of Model (C) are all 
SM singlets, i.e. $\chi^-$, $(\zeta_R^{++},\zeta_R^+)$, $E_{L,R}$, and 
$(\phi_R^+,\phi_R^0)$, so they will not affect the oblique parameters. 

\noindent \underline{\it Details of Model (C)}~:~\\
The charged leptons $e,\mu,\tau$ obtain masses through the heavy singlets 
$E_{1,2,3}$ in the $6 \times 6$ Dirac mass matrix linking 
$(\bar{e}_L,\bar{\mu}_L,\bar{\tau}_L;\bar{E}_{1L},\bar{E}_{2L},\bar{E}_{3L})$ 
to $(e_R,\mu_R,\tau_R;E_{1R},E_{2R},E_{3R})$:
\begin{equation}
{\cal M}_{eE} = \pmatrix{0 & {\cal M}_L \cr {\cal M}_R & {\cal M}_E},
\end{equation}
where ${\cal M}_{L,R}$ are $3 \times 3$ mass matrices proportional to 
$\langle \phi^0_{L,R} \rangle$ respectively.  Hence
\begin{equation}
{\cal M}_e = {\cal M}_L {\cal M}_E^{-1} {\cal M}_R.
\end{equation}

In the scalar sector, in addition to the $\Phi_{L,R}$ doublets, there are 
the charged singlet $\chi^-$ and the exotic doublet 
$\zeta = (\zeta_R^{++},\zeta_R^+)$.  From the structure of Fig.~3, it is 
clear that $\chi^-$ carries lepton number $L=+2$ and $\zeta$ carries $L=-2$. 
The most general scalar potential is then 
given by
\begin{eqnarray}
V &=& -\mu_L^2 \Phi_L^\dagger \Phi_L - \mu_R^2 \Phi_R^\dagger \Phi_R 
+ \mu_1^2 \chi^+ \chi^- + \mu_2^2 \zeta^\dagger \zeta + 
[\mu_3 \chi^- (\Phi_R^\dagger \zeta) + H.c.] \nonumber \\ 
&+& {1 \over 2} \lambda_L (\Phi_L^\dagger \Phi_L)^2 + {1 \over 2} \lambda_R 
(\Phi_R^\dagger \Phi_R)^2 + {1 \over 2} \lambda_\chi (\chi^+ \chi^-)^2 + 
{1 \over 2} \lambda_\zeta (\zeta^\dagger \zeta)^2 \nonumber \\ 
&+& \lambda_{LR} (\Phi_L^\dagger \Phi_L)(\Phi_R^\dagger \Phi_R) + 
\lambda_{L\chi} (\Phi_L^\dagger \Phi_L)(\chi^+ \chi^-) + 
\lambda_{L\zeta} (\Phi_L^\dagger \Phi_L)(\zeta^\dagger \zeta) \nonumber \\ 
&+& \lambda_{R\chi} (\Phi_R^\dagger \Phi_R)(\chi^+ \chi^-) + 
\lambda_{R\zeta} (\Phi_R^\dagger \Phi_R)(\zeta^\dagger \zeta) + 
\lambda_{\chi\zeta} (\chi^+ \chi^-)(\zeta^\dagger \zeta).
\end{eqnarray}
As the gauge symmetries $SU(2)_{L,R}$ get broken by 
$\langle \phi^0_{L,R} \rangle = v_{L,R}$, where
\begin{eqnarray}
\mu_L^2 &=& \lambda_L v_L^2 + \lambda_{LR} v_R^2, \\ 
\mu_R^2 &=& \lambda_R v_R^2 + \lambda_{LR} v_L^2, 
\end{eqnarray}
only $\sqrt{2} Re (\phi^0_{L,R}) = h_{L,R}$ become 
physical fields, with
\begin{equation}
m_{h_L}^2 = 2 \lambda_L v_L^2, ~~~ m_{h_R}^2 = 2 \lambda_R v_R^2.
\end{equation}
The other physical scalars are $\chi^-$, $\zeta_R^{++}$, 
and $\zeta_R^+$.  The $2 \times 2$ mass-squared matrix spanning 
$(\chi^\pm, \zeta_R^\pm)$ is of the form
\begin{equation}
{\cal M}^2_{\chi, \zeta} = \pmatrix{m_\chi^2 & \mu_3 v_R \cr \mu_3 v_R & 
m_{\zeta^{++}}^2},
\end{equation}
where
\begin{eqnarray}
m_\chi^2 &=& \mu_1^2 + \lambda_{L\chi} v_L^2 + \lambda_{R\chi} v_R^2, \\ 
m_{\zeta^{++}}^2 &=& \mu_2^2 + \lambda_{L\zeta} v_L^2 + 
\lambda_{R\zeta} v_R^2. 
\end{eqnarray}
Let the two mass eigenstates be 
\begin{equation}
h^\pm_1 = \chi^\pm \cos \theta - \zeta_R^\pm \sin \theta, ~~~ 
h^\pm_2 = \chi^\pm \sin \theta + \zeta_R^\pm \cos \theta, 
\end{equation}
with $m_1 < m_2$, then Eq.~(10) implies that $m_1 < m_{\zeta^{++}} < m_2$.

The decay of $\zeta_R^{++}$ is possible through the allowed coupling 
$\zeta_R^{++} l_{iR} E_{jR}$ and the conversion of $E_R$ to $l_R$ through 
the $6 \times 6$ mass matrix of Eq.~(4).  This is thus a two-body decay 
with final state $l_{iR} l_{jR}$, analogous to the Higgs triplet case 
for neutrino mass but with the opposite chirality.  Another possible decay 
is to $h^+_1$ and a (virtual) $W_R^+$ which becomes $l_i^+ \nu_i$ or 
$\bar{d}_i u_i$, with $h^+_1$ decaying subsequently to $l_j^+ \bar{\nu}_j$, as 
shown in Fig.~5.
\begin{figure}[htb]
\vspace*{-3.5cm}
\hspace*{-3cm}
\includegraphics[scale=1.0]{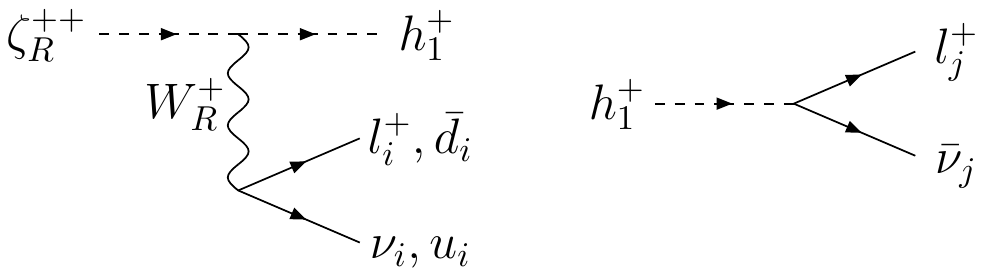}
\vspace*{-23.5cm}
\caption{Decay of the exotic scalar $\zeta_R^{++}$.}
\end{figure}
Whereas the two-body decay is suppressed by $l-E$ mixing, the three-body 
decay is suppressed by $\sin \theta$ of Eq.~(13).  If the latter is not 
small, then the amplitude of Fig.~5 may be significant or even dominant. 
Its signature is then $\zeta_R^{++} \to l^+_i l^+_j$ + missing energy, or 
$\zeta_R^{++} \to l^+$ + 2 jets + missing energy.  For $\zeta_R^{\pm \pm}$ 
pair production at the LHC, a distinctive final state of 4 charged leptons 
+ missing energy may be observed.  Future dedicated analyses based on the 
decay chain shown in Fig.~5 would be necessary to extract information on the 
existence of $\zeta_R^{++}$.

\noindent \underline{\it The Scalar Leptoquark Option}~:~\\
The particles in the loop of Fig.~3, i.e. $e_L,E_R,\chi^-,\zeta_R^+$, may 
be replaced by their color counterparts, i.e. 
$d_L,D_R,\chi^{-1/3},\zeta_R^{1/3}$.  In that case, the distinctive 
feature of the model is the exotic scalar leptoquark doublet 
$(\zeta_R^{4/3},\zeta_R^{1/3})$.  The singlet-doublet mixing of $\chi^{\pm1/3}$ 
with $\zeta_R^{\pm1/3}$ is of the same form as Eq.~(10).  Hence 
$\zeta_R^{-4/3}$ decays to $l_i d_j$ through $d-D$ mixing, as well as 
$W_R^- (\nu_i d_j - l_i u_j)$ in analogy to Fig.~5.  If the latter 
dominates, the signature is again not that of the conventional scalar 
leptoquark, and requires more detailed analysis of future LHC data to 
search for its existence.

\noindent \underline{\it Exchange of $W_L$ and $W_R$}~:~\\
A radiative Dirac neutrino mass is generated through $W_L-W_R$ mixing, 
so that $m_\nu/m_l$ has a natural lower bound~\cite{m87}. 
In the absence of a scalar bidoublet, $W_L$ and $W_R$ do not mix. 
However, once the $u$ and $d$ quarks obtain masses, mixing does 
occur~\cite{bem89,bh89} in one loop.  Together with the appearance 
of charged-lepton masses, Dirac neutrino masses are induced in two 
loops.  These are presumably subdominant effects to our one-loop 
Dirac neutrino masses.  In addition, the analog of the $2W$ exchange 
diagram~\cite{bm88} for Majorana neutrino mass in the SM also exists, 
as shown in Fig.~6.
\begin{figure}[htb]
\vspace*{-2cm}
\hspace*{1cm}
\includegraphics[scale=0.6]{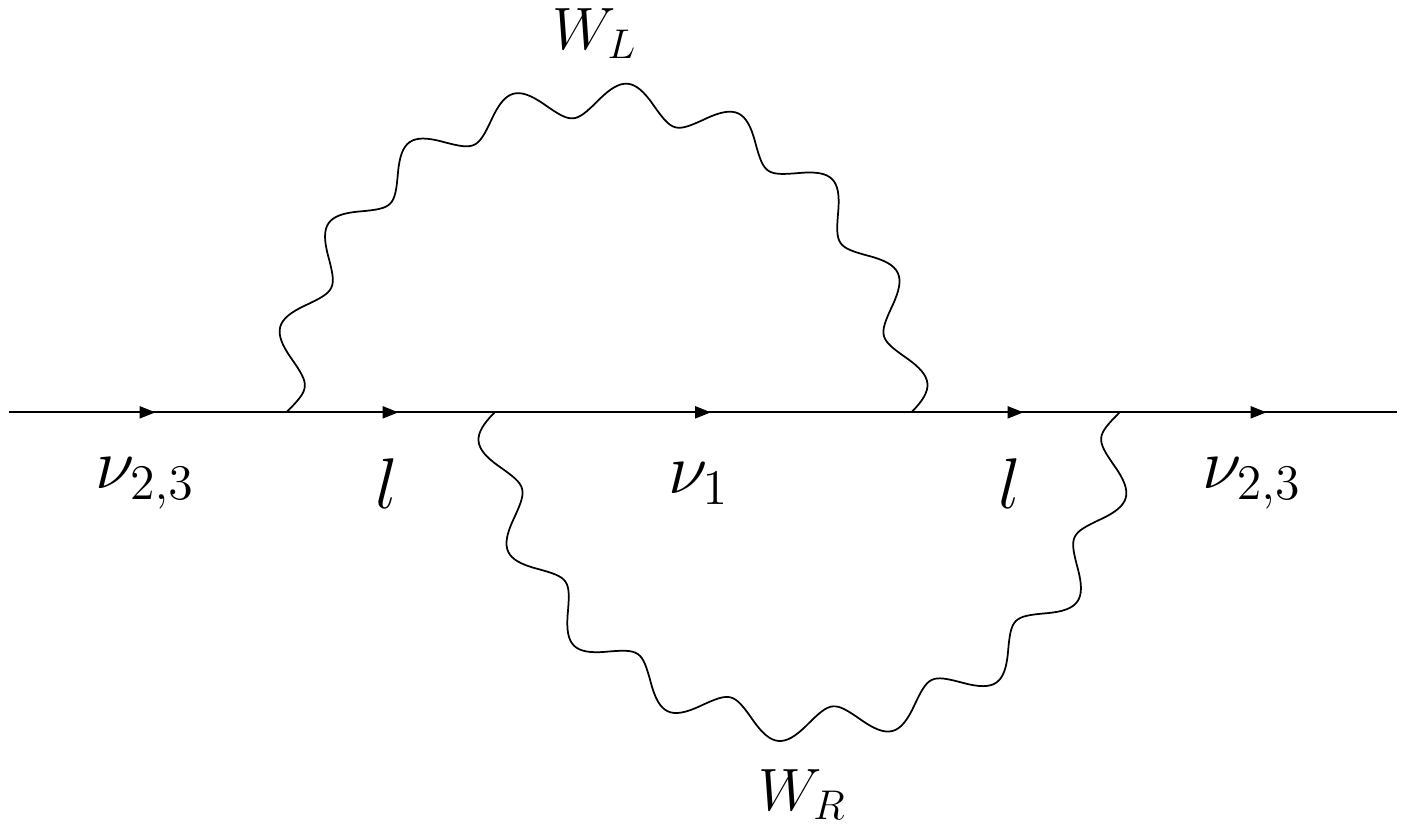}
\vspace*{-10.5cm}
\caption{Dirac neutrino mass from $W_{L,R}$ exchange.}
\end{figure}
This shows that even if only one neutrino, say $\nu_1$, picks up a 
Dirac mass from Eq.~(2), the other neutrinos $\nu_{2,3}$ will get 
nonzero radiative masses as well, although in practice they are 
numerically negligible.

\noindent \underline{\it Concluding Remarks}~:~\\
In the context of the well-known left-right gauge model of quarks and 
leptons, we study the case where the scalar bidoublet is absent.  
Assuming that quark and charged leptons obtain their masses through 
their corresponding heavy singlets using the Dirac seesaw mechanism, 
we consider how one-loop Dirac neutrino masses may be obtained. 
We identify four generic scenarios, two of which were proposed almost 
30 years ago.  We focus on one of the two new scenarios, with a new 
charged singlet scalar $\chi^-$ and an exotic $SU(2)_R$ scalar doublet 
$(\zeta_R^{++},\zeta_R^+)$.  We show how $\zeta_R^{++}$ differs from the 
well-known doubly charged scalar in the Higgs triplet model of neutrino 
mass, and discuss its distinctive decay signature at the LHC. 
We also point out that there could be a scalar leptoquark variant of 
this radiative mechanism using the exotic doublet 
$(\zeta_R^{4/3},\zeta_R^{1/3})$.

\noindent \underline{\it Acknowledgement}~:~\\
The work of E.M. was supported in part by the 
U.~S.~Department of Energy under Grant No. DE-SC0008541 and the work of 
U.S. was supported by a research grant associated with the J.~C.~Bose 
Fellowship, DST, India.


\bibliographystyle{unsrt}

\begin{thebibliography}{99}
\bibitem{w79} S. Weinberg, Phys. Rev. Lett. {\bf 43}, 1566 (1979).
\bibitem{m98} E. Ma, Phys. Rev. Lett. {\bf 81}, 1171 (1998).
\bibitem{mp17} E. Ma and O. Popov, Phys. Lett. {\bf B764}, 142 (2017).
\bibitem{bms03} B. Brahmachari, E. Ma, and U. Sarkar, Phys. Rev. Lett. 
{\bf 91}, 011801 (2003).
\bibitem{m88} R. N. Mohapatra, Phys. Lett. {\bf B201}, 517 (1988).
\bibitem{z80} A. Zee, Phys. Lett. {\bf B93}, 389 (1980).
\bibitem{m89} E. Ma, Phys. Rev. Lett. {\bf 63}, 1042 (1989).
\bibitem{m06} E. Ma, Phys. Rev. {\bf D73}, 077301 (2006).
\bibitem{bd17} D. Borah and A. Dasgupta, JCAP {\bf 1706}, 003 (2017).
\bibitem{ms98} E. Ma and U. Sarkar, Phys. Rev. Lett. {\bf 80}, 5716 (1998).
\bibitem{mrs00} E. Ma, M. Raidal, and U. Sarkar, Phys. Rev. Lett. {\bf 85}, 
3769 (2000).
\bibitem{atlas16} ATLAS Collaboration, Report No. ATLAS-CONF-2016-051.
\bibitem{bj17} K. S. Babu and S. Jana, Phys. Rev. {\bf D95}, 055020 (2017).
\bibitem{ds16} See for example D. Das and A. Santamaria, Phys. Rev. {\bf D94}, 
015015 (2016).
\bibitem{m87} E. Ma, Mod. Phys. Lett. {\bf A2}, 63 (1987).
\bibitem{bem89} K. S. Babu, D. Eichler, and R. N. Mohapatra, Phys. Lett. 
{\bf B226}, 347 (1989).
\bibitem{bh89} K. S. Babu and X.-G. He, Mod. Phys. Lett. {\bf A4}, 61 (1989).
\bibitem{bm88} K. S. Babu and E. Ma, Phys. Rev. Lett. {\bf 61}, 674 (1988).

\end{thebibliography}

\end{document}